\documentclass[12pt]{article}
\usepackage{graphicx}
\usepackage{hyperref}

\newcommand\pubnumber{NuPhys2015-Meagher}
\newcommand\pubdate{\today}

\def\Title#1{\begin{center} {\Large #1 } \end{center}}
\def\Author#1{\begin{center}{ \sc #1} \end{center}}
\def\Address#1{\begin{center}{ \it #1} \end{center}}

\newcommand\pubblock{\rightline{\begin{tabular}{l} \pubnumber\\
         \pubdate  \end{tabular}}}
\newenvironment{Abstract}{\begin{quotation}  }{\end{quotation}}
\newenvironment{Presented}{\begin{quotation} \begin{center} 
             PRESENTED AT\end{center}\bigskip 
      \begin{center}\begin{large}}{\end{large}\end{center} \end{quotation}}

\def\beq{\begin{equation}}
\def\eeq#1{\label{#1}\end{equation}}
\def\eeqn{\end{equation}}

\def\beqa{\begin{eqnarray}}
\def\eeqa#1{\label{#1}\end{eqnarray}}
\def\eeqan{\end{eqnarray}}

\let\bar=\overbar

\def\etal{{\it et al.}}

\def\Dslash{\not{\hbox{\kern-4pt $D$}}}
\def\dslash{\not{\hbox{\kern-2pt $\del$}}}

\def\msb{{\bar{\ssstyle M \kern -1pt S}}}

\newcommand{\degree}{^{\circ}}
\newcommand{\arXiv}[1]{\href{https://arxiv.org/abs/#1}{\texttt{arXiv:#1}}}
\newcommand{\email}[1]{\href{mailto:#1}{\nolinkurl{#1}}}

\let\OLDthebibliography\thebibliography
\renewcommand\thebibliography[1]{
  \OLDthebibliography{#1}
  \setlength{\parskip}{0ex}
  \setlength{\itemsep}{0.4ex}
}

\begin{document}
\begin{titlepage}
\pubblock

\vfill
\Title{Neutrino Astronomy with IceCube and Beyond}
\vfill
\Author{ Kevin J. Meagher \\
  on behalf of the IceCube Collaboration}
\Address{
  Universit{\'e} Libre de Bruxelles, Science Faculty CP230,\\
  B1050 Bruxelles, Belgium \\
  \rm{email: \email{kmeagher@ulb.ac.be}\\
  website: \url{http://icecube.wisc.edu}}
  }
\vfill
\begin{Abstract}
  The IceCube Neutrino Observatory is a cubic kilometer neutrino telescope
  located at the geographic South Pole. Cherenkov radiation
  emitted by charged secondary particles from neutrino interactions is
  observed by IceCube using an array of 5160 photomultiplier tubes embedded
  between a depth of 1.5\,km to 2.5\,km in the Antarctic glacial ice. The detection
  of astrophysical neutrinos is a primary goal of IceCube and has
  now been realized with the discovery of a diffuse, high-energy flux consisting
  of neutrino events from tens of TeV up to several PeV. Many
  analyses have been performed to identify the source of these neutrinos, including
  correlations with active galactic nuclei, gamma-ray bursts, and the Galactic
  plane. IceCube also conducts multi-messenger campaigns to alert other observatories
  of possible neutrino transients in real time. However, the source of
  these neutrinos remains elusive as no corresponding electromagnetic
  counterparts have been identified. This proceeding will give an overview of
  the detection principles of IceCube, the properties of the observed astrophysical
  neutrinos, the search for corresponding sources (including
  real-time searches), and plans for a next-generation neutrino detector,
  IceCube--Gen2.
\end{Abstract}
\vfill
\begin{Presented}
  NuPhys2016, Prospects in Neutrino Physics\\
  Barbican Centre, London, UK,  December 12--14, 2016
\end{Presented}
\vfill
\end{titlepage}
\def\thefootnote{\fnsymbol{footnote}}
\setcounter{footnote}{0}

\section{Neutrino Astronomy}
From radio waves to gamma rays, electromagnetic radiation has been the source of a wealth of information about the universe.
Unfortunately, photons with energies above 1\,TeV are absorbed by extra-galactic background light, making it difficult to detect sources beyond a redshift of 0.1 above this energy\cite{opacity}.
In order to study the universe above this cut-off we need to find an alternative to photons.
Cosmic rays tell us that charged particles are accelerated by astrophysical objects up to at least $10^{20}$\,eV, but since charged particles are deflected by magnetic fields, the origin of these particles still remains unclear.
Since, aside from gravity, neutrinos interact solely via the weak force, they can traverse the universe completely unimpeded and therefore hold the potential to open a new window on astronomy.

\section{The IceCube Neutrino Observatory}
Neutrinos' small cross section, the same property that allows them to arrive at Earth unimpeded, also makes them difficult to detect.
Observing neutrinos requires a large target mass to make up for the small cross section.
In addition, the medium must be transparent in order to observe the light from the secondary particles.
The IceCube Neutrino Observatory was built in the Antarctic ice sheet at the South Pole Station.

The fundamental unit of IceCube is the digital optical module (DOM).
Each DOM contains a 25\,cm photomultiplier tube, high voltage power supply, and digitization and communication electronics.
DOMs are aligned on vertical structures called strings, with 60 DOMs per string spaced vertically by 17\,m between a depth of 1450\,m and 2450\,m.
There are 86 strings for a total of 5160 DOMs.
The strings form a triangular grid with a spacing of 125\,m, except for 8 strings arrayed in the center to form a denser formation referred to as DeepCore.

There are three main event selections used for neutrino astronomy: muon tracks, cascades, and high energy starting-events (HESE).
Muon tracks have good angular resolution, ${\sim}0.7^\circ$ for energies above 10\,TeV, but not all of the energy is deposited in the detector and so have comparatively poor energy resolution.
With cascades, all of the energy is deposited near the vertex. These events have much better energy resolution than tracks, but at the cost of relatively poor angular resolution.
The HESE selection observes events which start in the detector volume, by only selecting events where the initial light occurs on DOMs within the interior of the detector, and vetoing events which start near the edge.
Although the events in this selection are either tracks or cascades, from an analysis point of view HESE is a separate event selection.

\section{Observation of High-Energy Neutrinos}
\begin{figure}[tb]
  \begin{center}
    \includegraphics[width=0.48\textwidth]{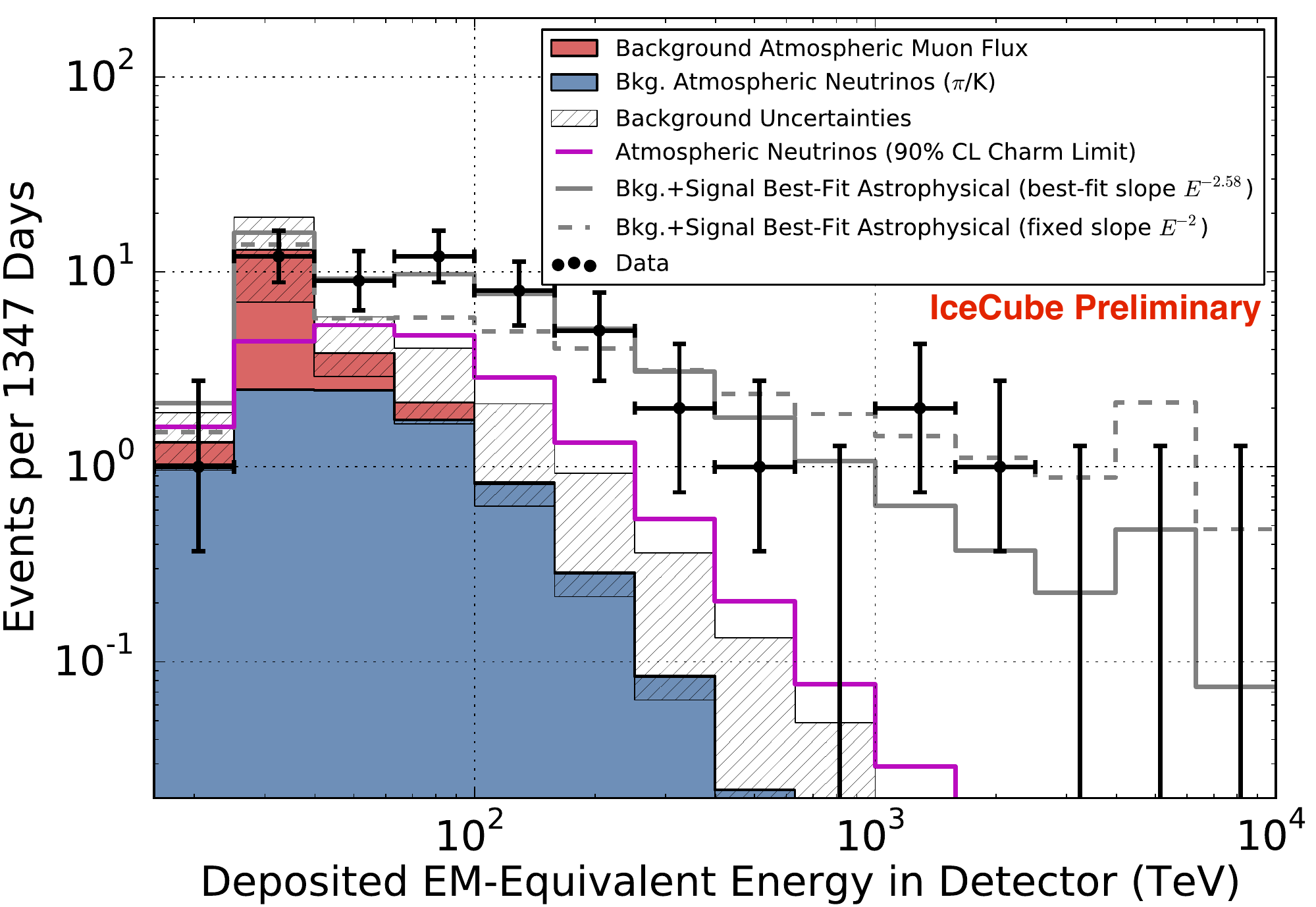}%
    \includegraphics[width=0.52\textwidth]{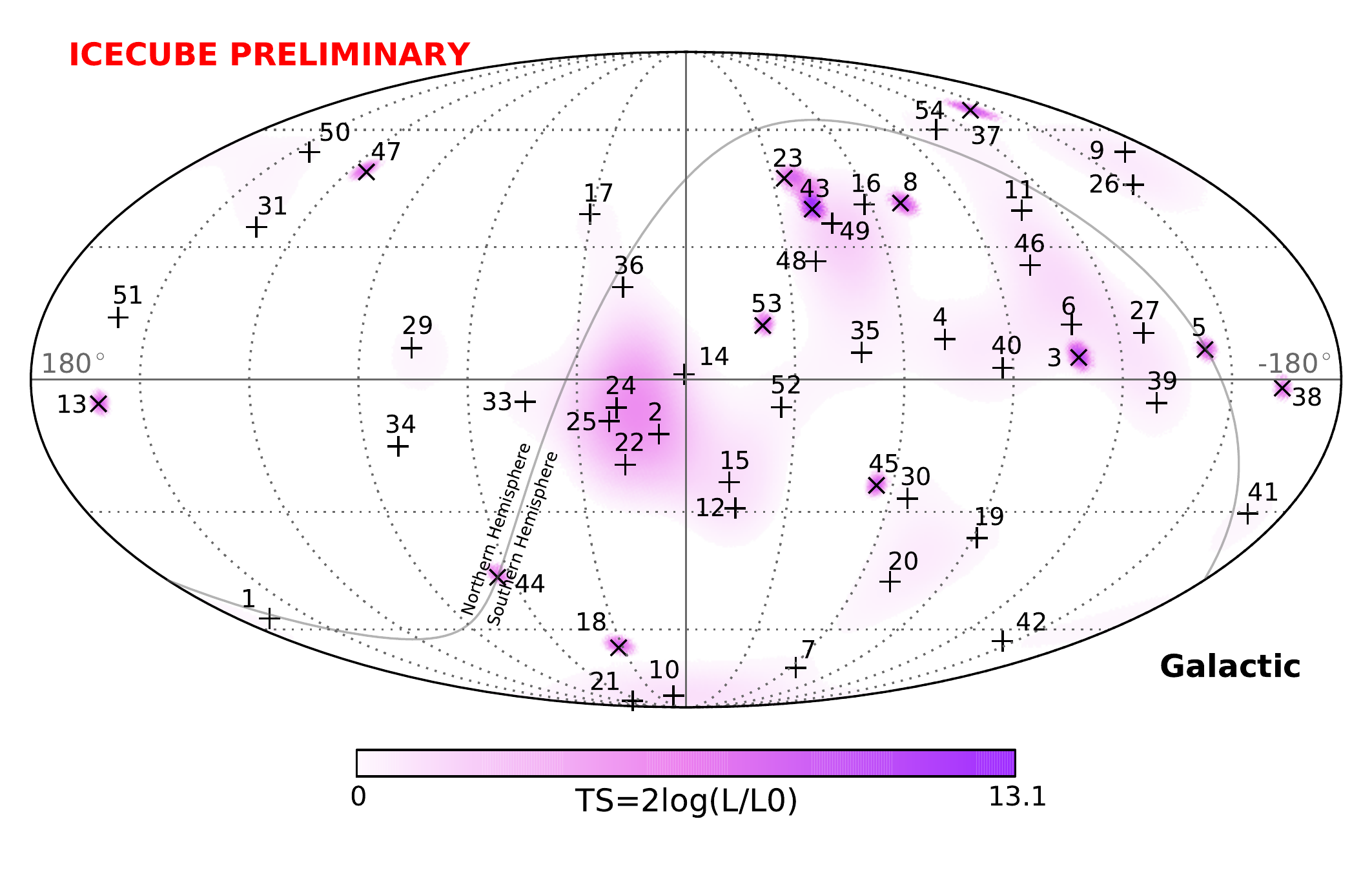}     
    \caption{
      \textbf{Left:}
      Deposited energy spectrum of the high energy starting-event (HESE) sample\cite{HESE}.
      The hashed region shows uncertainties on the sum of
      all backgrounds. Muons are computed from simulation
      to overcome statistical limitations in our background measurement
      and scaled to match the total measured background
      rate. Atmospheric neutrinos and uncertainties thereon are derived
      from previous measurements of the $\pi$/K component, and upper limits on the charm
      component of the atmospheric $\nu_\mu$ spectrum.
      \textbf{Right:}
      Arrival directions of the same sample in Galactic coordinates.
      Shower-like events are marked with ``+'' and those containing tracks with ``$\times$''.
      The color scale shows the test statistic for the point-source clustering test at each location.
      No significant clustering was found.
    }
    \label{fig:hese}
  \end{center}
\end{figure}
Using the HESE sample, an analysis performed on 4 years of data found 54 neutrino candidate events with a statistical significance of $6.5\sigma$ \cite{HESE}.
In order to describe the data, a maximum likelihood, forward-folding fit of all components (atmospheric muons, atmospheric neutrinos from $\pi$/K decay, atmospheric neutrinos from charm decay and an astrophysical flux assuming a 1:1:1 flavor ratio) was performed on the energy spectrum.
The result of the fit, shown in Figure \ref{fig:hese}\,(left), is $\mathrm{d}N/\mathrm{d}E = (2.2\pm0.7) \times 10^{-18}\cdot(E/100\,\mathrm{TeV})^{-2.58\pm0.25}\mathrm{GeV^{-1}cm^{-2}s^{-1}sr^{-1}}$.
A maximum likelihood clustering method was used to look for any neutrino point sources in this sample. 
This test, shown in Figure \ref{fig:hese}\,(right), did not yield significant evidence of clustering, with p-values of 44\% and 58\% for the shower-only and the all-events tests, respectively.
A test for Galactic plane clustering was also performed.
Assuming a Galactic width of $2.5\degree$ around the plane
resulted in a p-value of 7\% and a variable Galactic width scan resulted in a p-value 2.5\% (both p-values are trials corrected.)

\begin{figure}[tb]
  \begin{center}
    \includegraphics[width=0.56\textwidth]{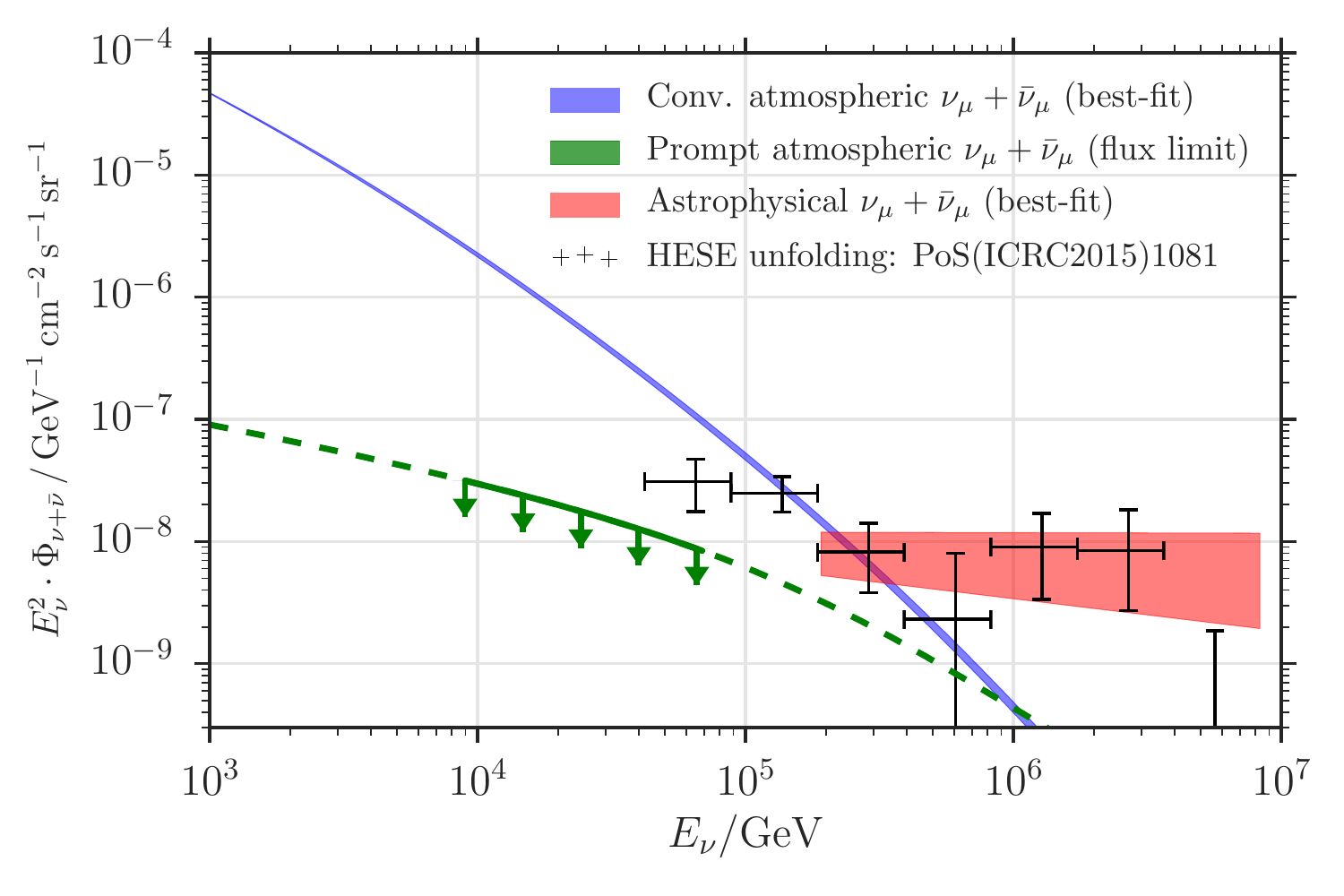}%
    \includegraphics[width=0.44\textwidth]{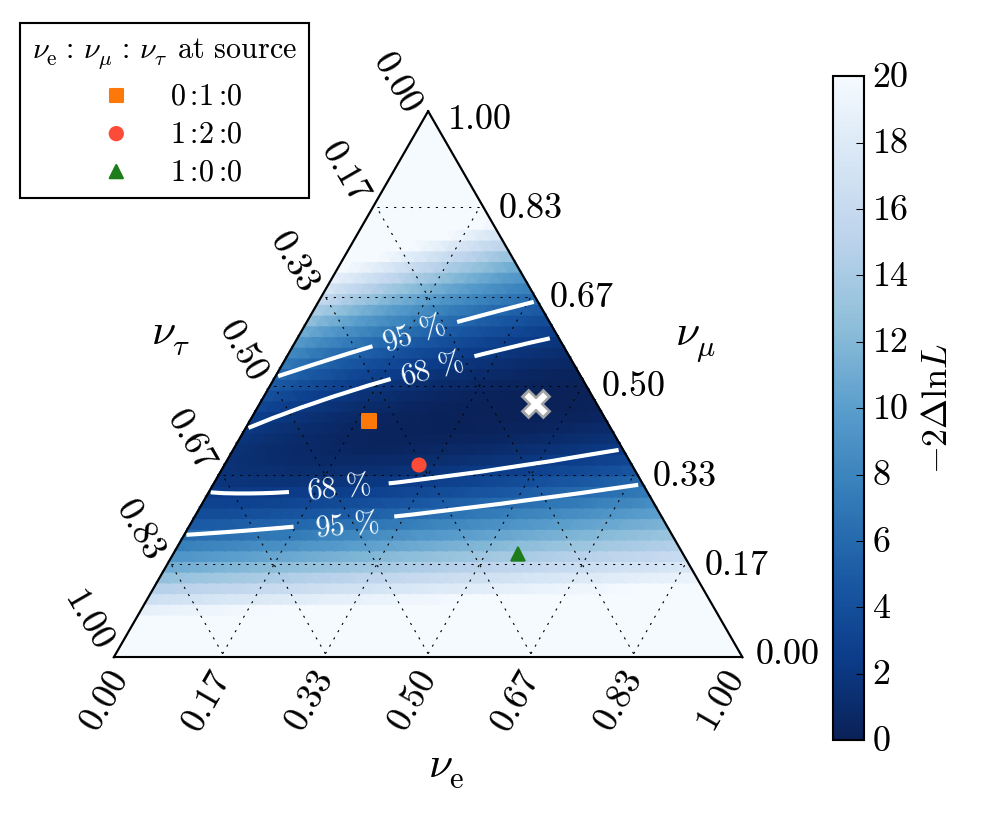} 
    \caption{
      \textbf{Left:}
      Best-fit neutrino spectra for the unbroken power law model using muon neutrino events.
      The conventional and astrophysical neutrino fluxes are represented by shaded regions indicating one sigma error
      on the measured spectrum, whereas the solid line represents the upper
      limit on the prompt neutrino model in \cite{charm}.
      The horizontal width of the astrophysical shaded region denotes the range of neutrino energies
      which contribute 90\% to the total likelihood ratio between
      the best-fit and the conventional atmospheric-only hypothesis. The
      crosses show the unfolded spectrum of the high-energy sample discussed above.
      \textbf{Right:}
      The results of the profile likelihood scan of the flavor composition at Earth.
      Each point in the triangle corresponds to a ratio $\nu_e:\nu_\mu:\nu_\tau$ as measured on Earth.
      The individual contributions are read off the three sides of the triangle.
      The best-fit composition is marked with ``$\times$''.
      68\% and 95\% confidence regions are indicated.
      The ratios corresponding to three flavor composition scenarios at the sources of the neutrinos
      are marked by the square for pion-decay (0:1:0), circle for muon-damped (1:2:0), and triangle for neutron-beam (1:0:0) sources respectively.
    }
    \label{fig:diffuse}
  \end{center}
\end{figure}
A separate diffuse spectral analysis was performed using six years of data with the muon track event sample \cite{diffuse}.
At neutrino interaction energies between 191\,TeV and 8.3\,PeV an astrophysical contribution was observed with a significance of $5.6\sigma$.
As shown in Figure \ref{fig:diffuse}\,(left), the data were well described by a power law:  $\mathrm{d}N/\mathrm{d}E = (0.90^{+0.30}_{-0.27}) \times 10^{-18}\cdot(E/100\,\mathrm{TeV})^{-2.13 \pm 0.13}\,\mathrm{GeV^{-1}cm^{-2}s^{-1}sr^{-1}}$.

The ratio of different neutrino flavors can give important clues to acceleration mechanisms of the source.
In \cite{diffuse} we also performed a measurement of the flavor composition of the astrophysical neutrino flux, in which the normalizations of all three flavors were allowed to vary independently.
The results, shown in Figure \ref{fig:diffuse}\,(right), are consistent with pion-decay sources and muon-damped sources but disfavor neutron-beam sources with a significance of $3.6\sigma$.

In the cascade event sample, in an analysis of the first two years of data a total of 172 events were observed with energies between 10\,TeV and 1\,PeV \cite{cascade}.
The astrophysical component is also well described by a power law: $\mathrm{d}N/\mathrm{d}E = (2.3^{+0.7}_{-0.6}) \times 10^{-18}\cdot(E/100\,\mathrm{TeV})^{-2.67 \pm 0.13}\,\mathrm{GeV^{-1}cm^{-2}s^{-1}sr^{-1}}$.
The background-only hypothesis is rejected with a significance of $4.7\sigma$.

\begin{figure}[tb]
  \begin{center}
    \includegraphics[width=0.5\textwidth]{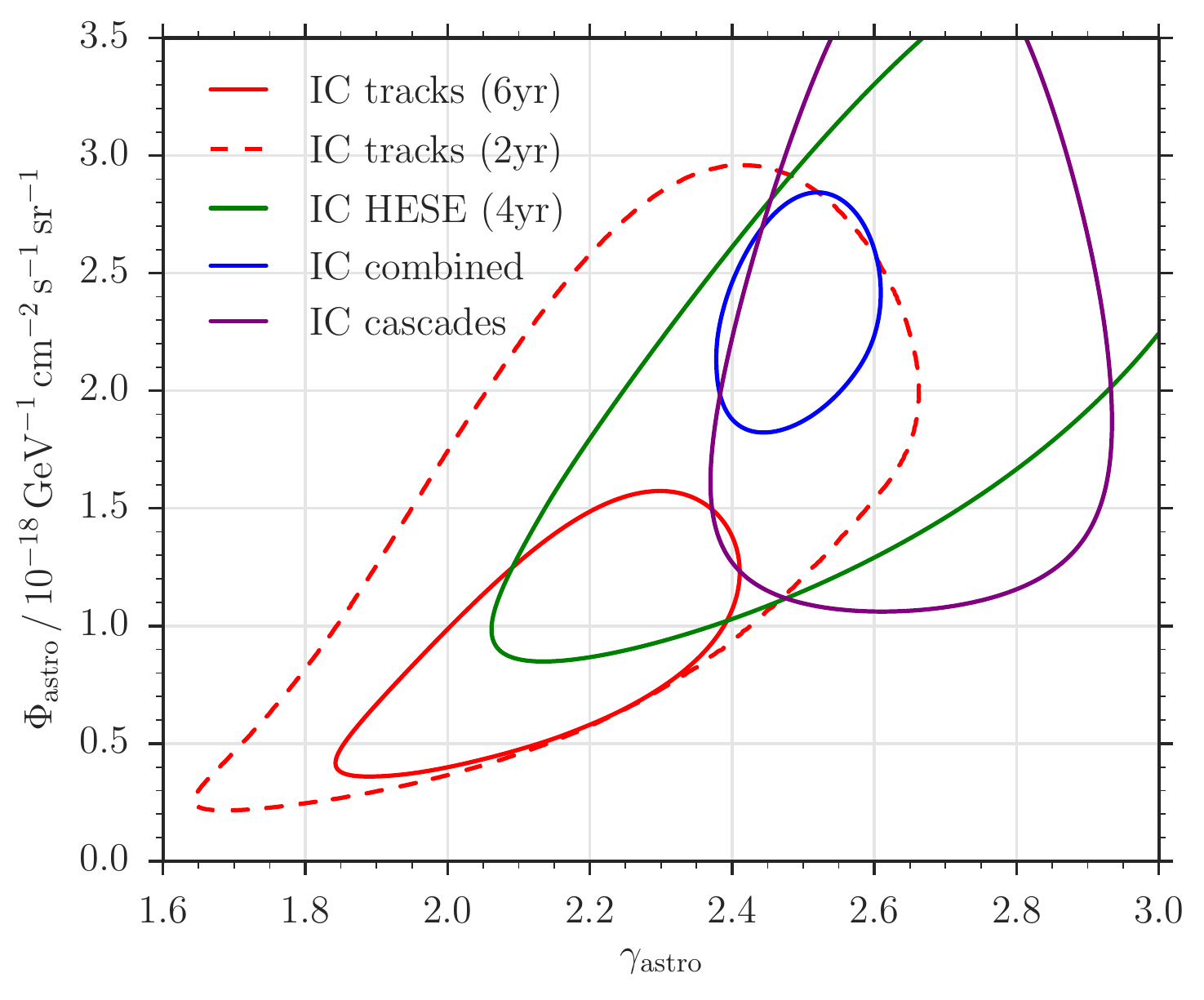}%
    \caption{
      Results of different IceCube analyses measuring the astrophysical flux parameters $\Phi_{astro}$ and $\gamma_{astro}$.
      The contour lines are at 90\% confidence
      The result of the muon track diffuse analysis \cite{diffuse} is shown by the bottom-left solid contour line.
      The contour obtained by the previous measurement using the same method but using 2 years of data is the dashed line.
      The results of the HESE analysis \cite{HESE}, cascade sample \cite{cascade}, and the combined analysis \cite{diffuse} are also shown.
    }
    \label{fig:fit}
  \end{center}
\end{figure}
The results of these analyses along with the results of three other diffuse analyses were combined into a global spectral analysis \cite{diffuse}.
Assuming the astrophysical neutrino flux to be isotropic and to consist of equal flavors at Earth, the all-flavor spectrum with neutrino energies between 25\,TeV and 2.8\,PeV is well described by an unbroken power law with a best-fit spectral index $−2.50\pm0.09$ and a flux at 100\,TeV of ${6.7}_{-1.2}^{+1.1}\cdot10^{-18}\,\mathrm{GeV^{-1}\,s^{-1}\,sr^{-1}\,cm^{-2}}$. Note that this flux is the sum of all three neutrino flavors, wheras the numbers quoted earlier in this section were per flavor fluxes.
The results of the combined sample spectral fit along with the previously mentioned analyses are shown in Figure \ref{fig:fit}.
Slight tension is seen between the different analyses.
Since the analyses which are more sensitive to higher energy neutrinos also have a greater sensitivity in the Northern Hemisphere, this tension may indicate either a spectral hardening at high energies or that the sources in the Northern Hemisphere have a harder spectrum than their southern counterparts. 

\section{The Search for Astrophysical Sources}
\begin{figure}[tb]
  \begin{center}
    \includegraphics[width=0.5\textwidth]{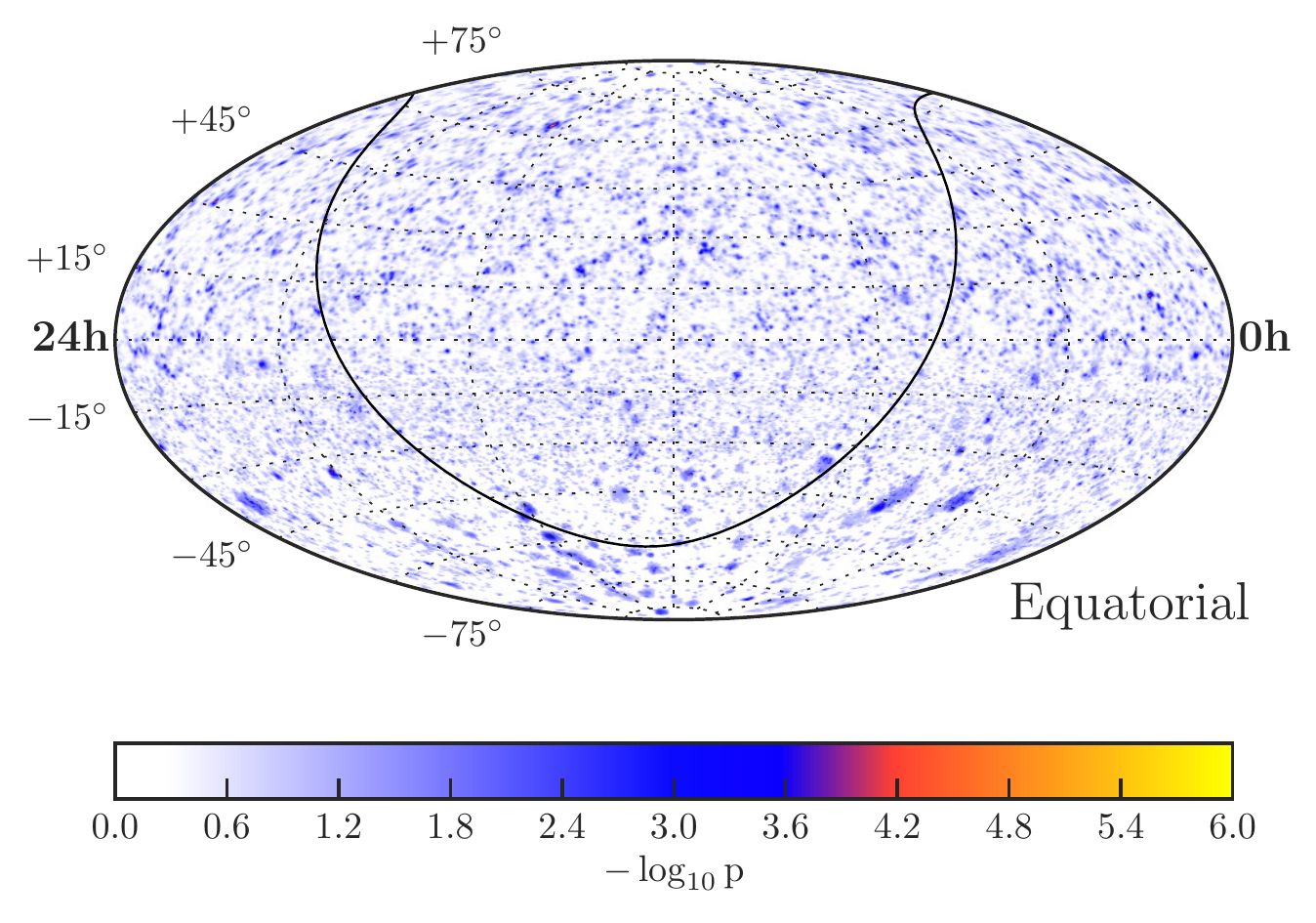}%
    \includegraphics[width=0.5\textwidth]{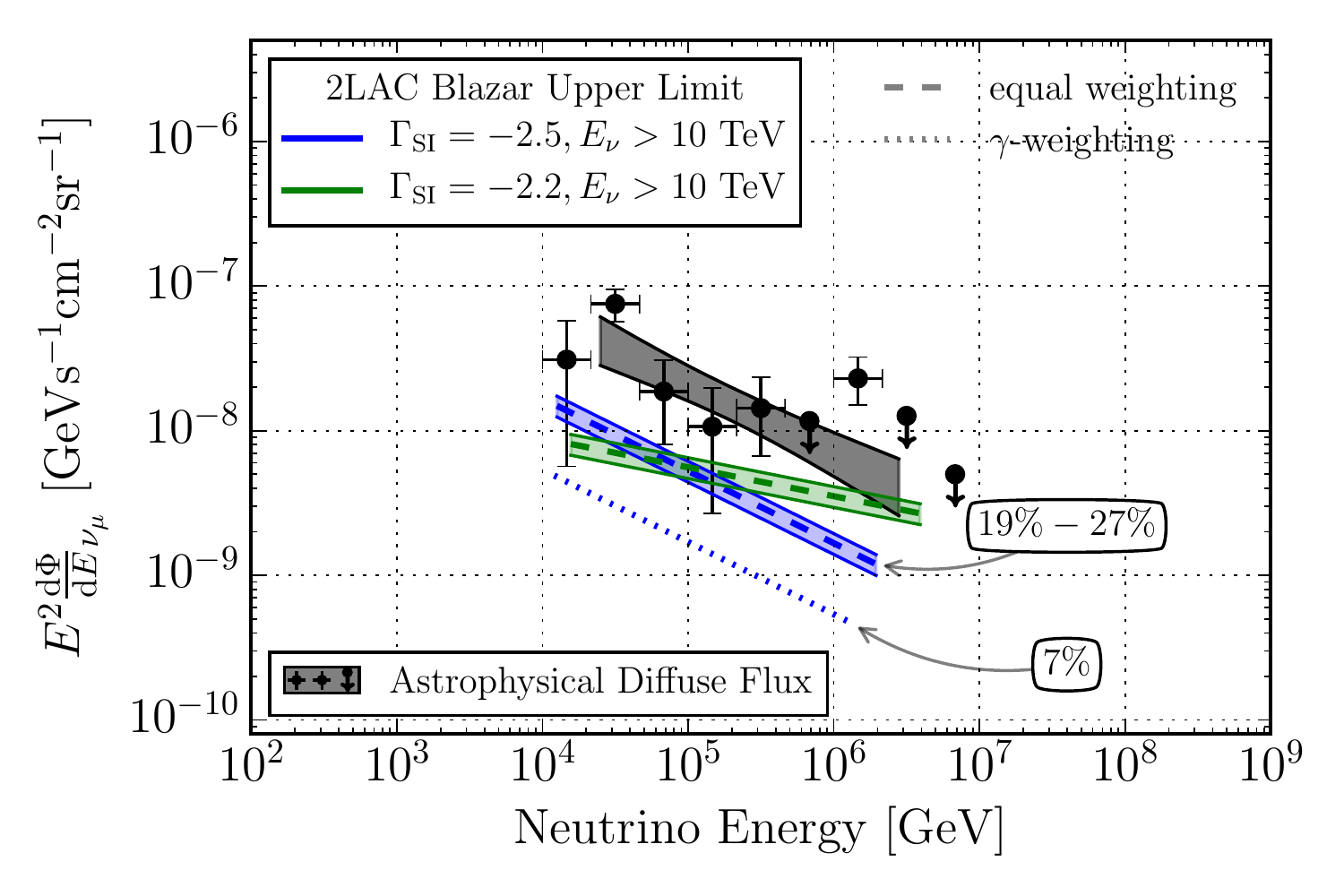} 
    \caption{
      \textbf{Left:}
      Pre-trial significance map of the all-sky point source scan \cite{ps}.
      The color indicates the negative logarithm of the pre-trial p-value assuming no clustering as the null hypothesis.
      Shown in Equatorial (J2000) coordinates, a line indicates the Galactic plane.
      \textbf{Right:}
      Results of the stacked blazar analysis.
      Neutrino flux upper limits from \cite{blazar} are shown in colors compared to the diffuse best fit from \cite{diffuse} shown in black.
      Shown in blue are two separate signal weighting schemes for an $E^{-2.5}$ energy spectrum:
      equal weighting (dashed line) where blazars are considered to contribute
      equally to the neutrino flux, and weighting by blazars' observed gamma-ray luminosity (dotted line).
      The equal-weighting upper limit for a flux with a harder spectral index of −2.2 is shown in green.
      Percentages denote the upper limit on the fraction of the integral astrophysical flux.
    }
    \label{fig:ps}
  \end{center}
\end{figure}
To identify the source of the neutrino populations described in the previous section, many analyses have been performed.
To date none of them has identified any association with known or unknown astrophysical sources.
In seven years of data, from 2008--2015, using an unbinned maximum-likelihood search for local clustering in the muon sample, no significant clustering of neutrinos above background expectation was observed \cite{ps}.
The map generated by this analysis is shown in Figure \ref{fig:ps}\,(left).
The negative result of this analysis excludes point sources with a flux above $E^{2}\mathrm{d}\Phi/\mathrm{d}E=10^{-12}\:\,\mathrm{TeV\,cm^{-2}\,s^{-1}}$.

Blazars have been proposed as a possible source of high-energy neutrinos\cite{blazar1}.
To investigate this a stacked analysis was performed with blazars from the 2nd Fermi-LAT AGN catalog (2LAC) \cite{blazar}.
No significant excess is observed, constraining the total population of 2LAC blazars to contributing 27\% or less of the observed astrophysical neutrino flux, assuming equipartition of neutrino flavors at Earth and the currently favored power law spectral index for the neutrino flux of 2.5.
As shown in Figure \ref{fig:ps}\,(right), the 2LAC blazars (and sub-populations) are excluded as being the dominant sources of the observed neutrinos up to a spectral index as hard as 2.2.

\begin{figure}[tb]
  \begin{center}
    \includegraphics[width=0.41\textwidth]{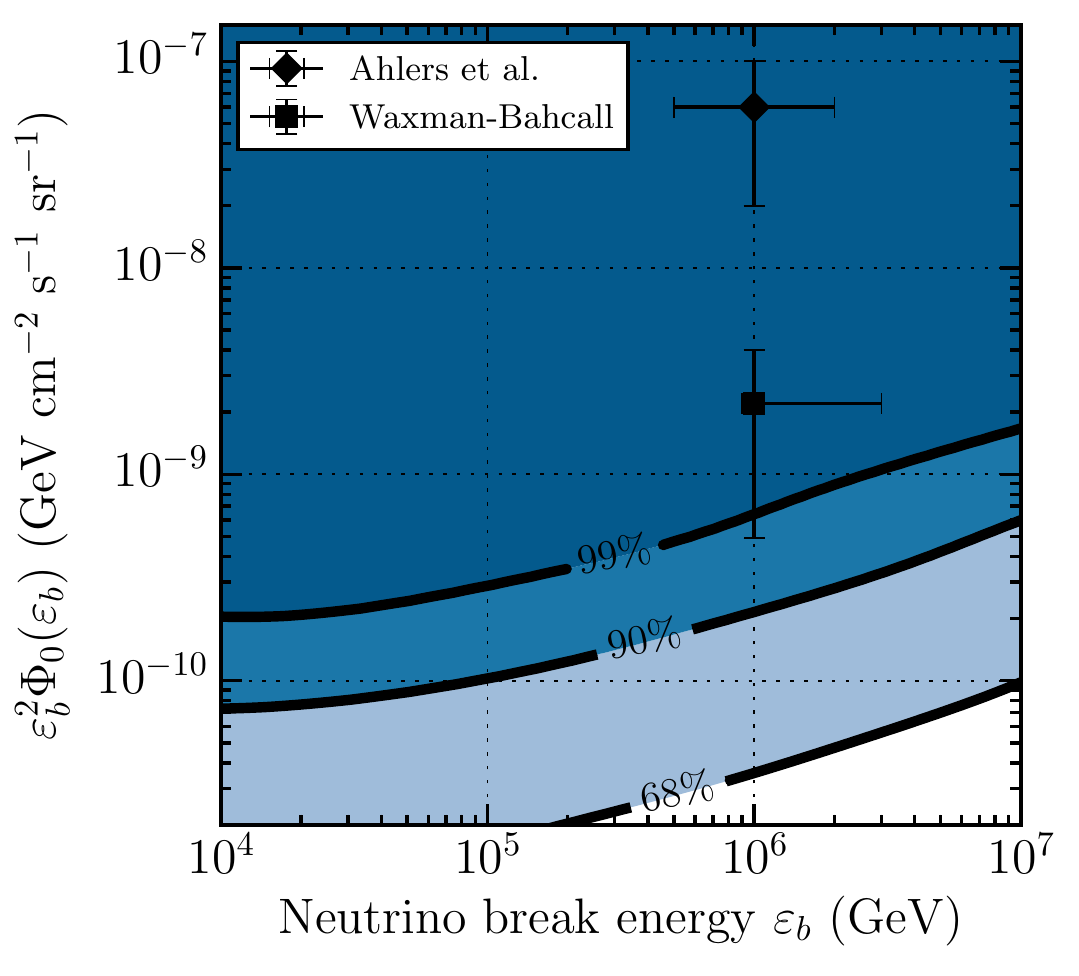}%
    \includegraphics[width=0.59\textwidth]{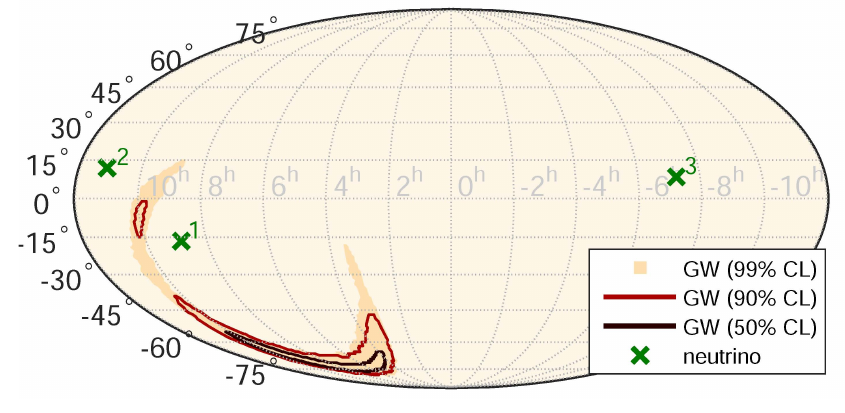} 
    \caption{
      \textbf{Left:}
      Constraint on generic doubly broken power law neutrino flux models as a function of
      first break energy $\varepsilon_b$ and normalization $\Phi_0$.
      The Ahlers \etal model \cite{Ahlers2011} assumes that
      only neutrons escape from the GRB fireball to contribute to the UHECR flux. The Waxman \&
      Bahcall model \cite{waxman-bahcall}, which allows all protons to escape the fireball, has been updated to account
      for more recent measurements of the UHECR flux \cite{Katz2009} and typical gamma-ray break
      energy \cite{Goldstein}.
      Exclusion contours, calculated from a combination of 5 years of muon track data with
      3 years of cascades\cite{grb}, are shown. 
      \textbf{Right:}
      Gravitational wave sky map in equatorial coordinates, showing
      the reconstructed probability density contours of the GW
      event at 50\%, 90\% and 99\% CL, and the reconstructed directions
      of high-energy neutrino candidates detected by IceCube
      (crosses) during a $\pm500$\,s time window around gravitational wave event GW150914.
      The neutrino directional uncertainties are $<\,1^{\circ}$ and are not shown.
      Gravitational wave shading indicates the reconstructed probability
      density of the gravitational wave event, darker regions corresponding to
      higher probability.
    }
    \label{fig:grb-exclusion}
  \end{center}
\end{figure}
Another astrophysical source considered to be a likely source of neutrinos are gamma-ray bursts (GRBs).
An analysis incorporating 5 years of muon track events and 1172 observed GRBs found no correlation more significant than expected from background \cite{grb}.
The limits on the neutrino flux set by this analysis (see Figure \ref{fig:grb-exclusion},left) disfavor much of the parameter space for the theories on neutrino emission from GRBs.
This analysis finds that no more than 1\% of the observed astrophysical neutrino flux consists of prompt emission from GRBs that are observable by existing satellites.

Another considered source was the first gravitational wave transient GW150914 observed by the Advanced LIGO detectors on Sept. 14th, 2015.
The analysis was performed by looking for neutrino candidates within 500\,s of the gravitational wave event.
As shown in Figure \ref{fig:grb-exclusion}\,(right) and consistent with background, three events were observed within this time window, none of them within the region triangulated by LIGO \cite{ligo}.

In order to alert other astronomers about possible neutrino transient events, the IceCube collaboration has developed several real-time alert programs.
The neutrino data are processed in real time at the South Pole Station and the most interesting neutrino events are selected to trigger observations with optical and X-ray telescopes aiming for the detection of an electromagnetic counterpart such as a GRB afterglow or a rising supernova light curve. The program is capable of triggering follow-up observations in less than a minute.
The optical follow-up program \cite{ofu} has been sending such alerts to optical telescopes since 2008 and to X-ray telescopes since 2009.
The gamma-ray follow-up program has been running since 2012 \cite{gfu}, sending triggers to the MAGIC and VERITAS gamma-ray telescopes.
This program focuses on blazar flares by monitoring a predefined list of known blazars and looks for excesses of neutrino events on timescales of up to three weeks.

\section{Future Upgrade}
\begin{figure}[tb]
  \begin{center}
    \includegraphics[width=3.4in]{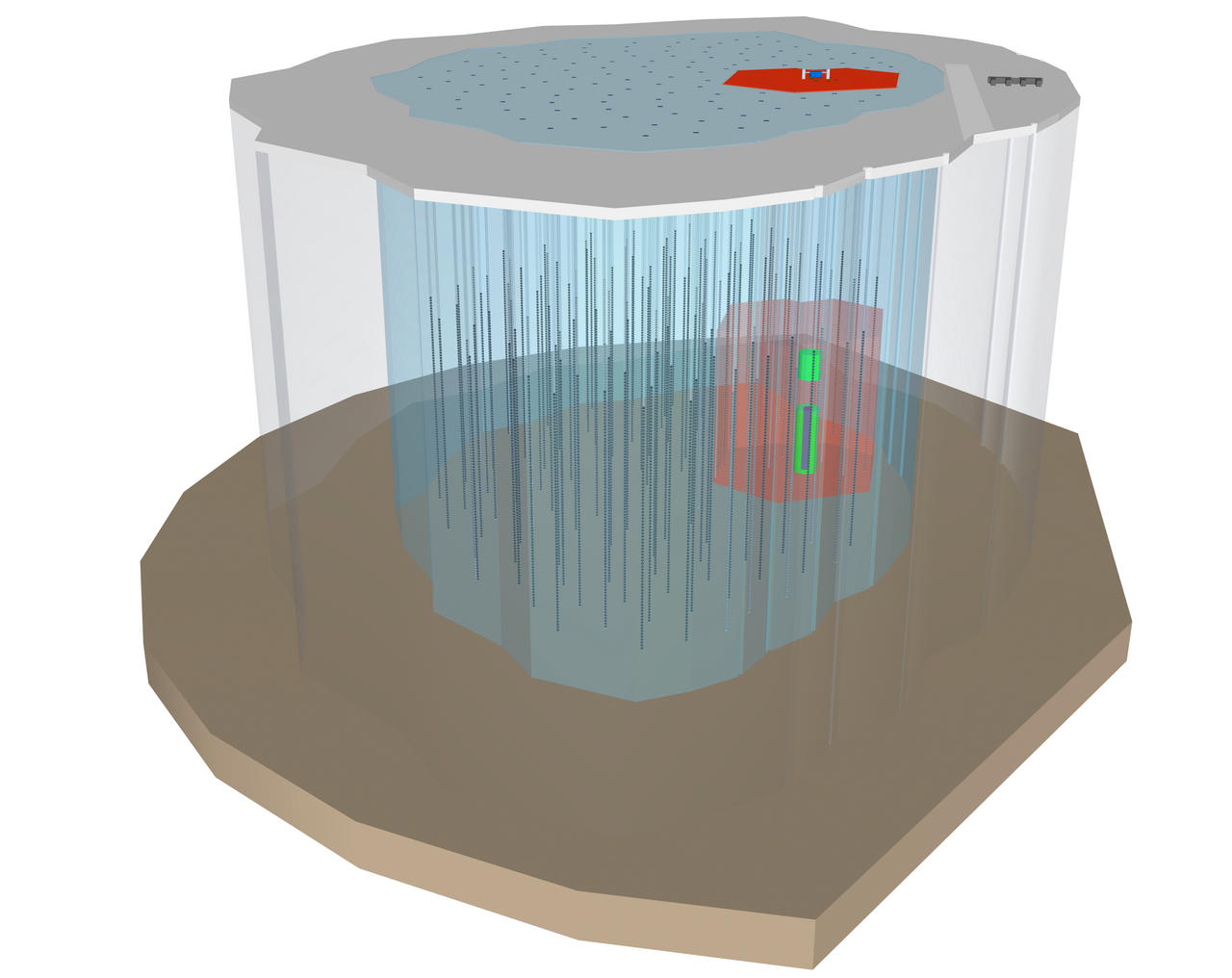} 
    \caption{
      Conceptual schematic of the proposed high-energy extension to the IceCube detector.
      The current IceCube is shown in red with DeepCore in green.
      120 additional string are added to increase the instrumented volume to ${\sim}10\,\mathrm{km}^3$.
      A veto detector, potentially comprised of scintillator detectors, is also envisioned at the surface.
      }
    \label{fig:schmatic}
  \end{center}
\end{figure}
Although IceCube has positively identified neutrinos of astrophysical origin, the ability of IceCube to be an efficient tool for neutrino astronomy over the next decade is limited by the modest number of cosmic neutrinos measured, even with a cubic kilometer array.
Design studies to increase IceCube's sensitivity with additional strings outside the current volume are currently underway \cite{gen2}.
This section will describe this effort, referred to as the IceCube--Gen2 High-Energy Array.
The design, shown in Figure \ref{fig:schmatic}, seeks to increase the instrumented volume to $\sim10\,\mathrm{km}^3$.
The high-energy array is proposed to complement the high-density, low-energy sub-array known as PINGU \cite{pingu}.
PINGU targets precision measurements of the atmospheric neutrino oscillation parameters and the determination of the neutrino mass hierarchy.

\begin{figure}[tb]
  \begin{center}
    \includegraphics[width=\textwidth]{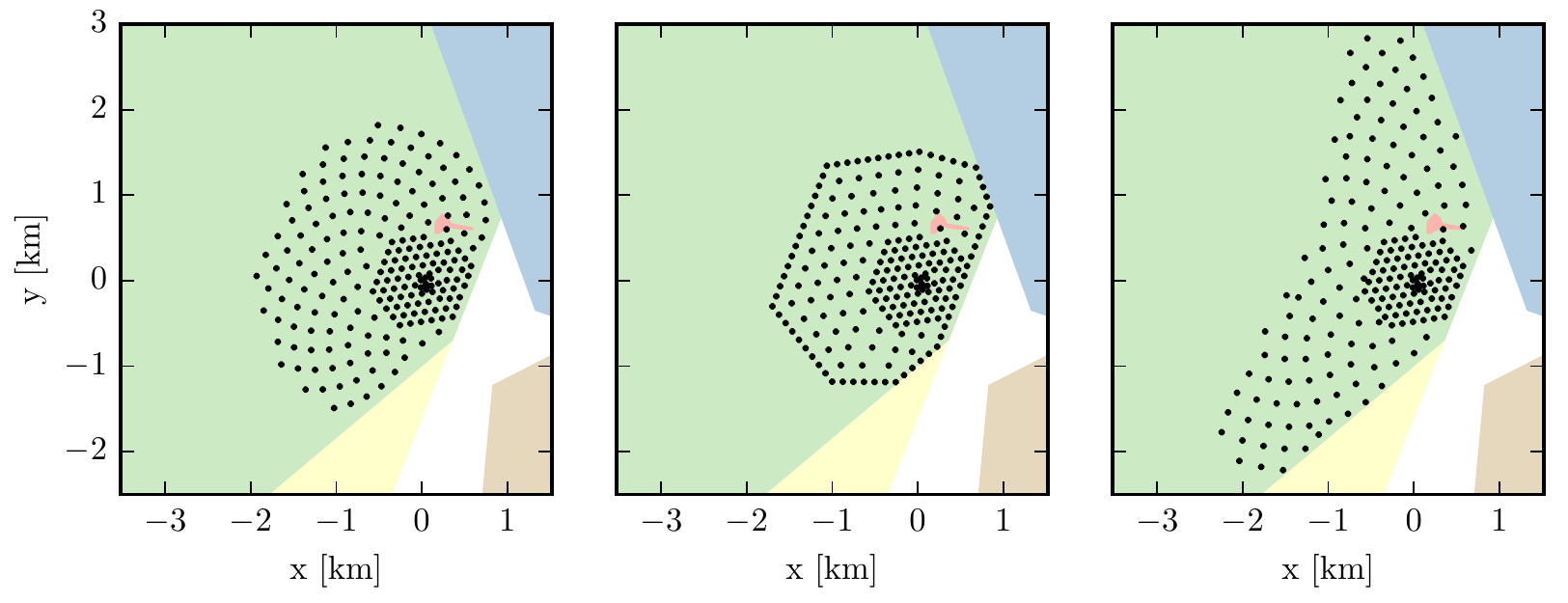} 
    \caption{
      Example benchmark detector string layouts under study for the high energy extension to IceCube.
      Each expands about IceCube by adding 120 strings constrained to the South Pole Dark Sector (shaded in light green).
      Other management areas of the South Pole Station where construction may not be permitted are also shown: the Downwind Sector in yellow, the Clean Air sector in blue, the Quiet Sector in beige, and the Old Station in pink.
      For the left panel, uniform string spacing of 240\,m is shown.
      The central panel represents a string layout with a denser edge weighting for improved veto efficiency.
      The right panel shows the so-called banana geometry which seeks to create a very long detector for certain muon tracks.
    }
    \label{fig:gen2-geometries}
  \end{center}
\end{figure}
The optical properties of deep Antarctic ice allow string spacing to be increased to 300\,m for energies exceeding 10\,TeV.
Since angular resolution for muon tracks is proportional to the length of the lever arm, by increasing the size of the detector, the angular resolution will also be improved, further improving point-source sensitivity.
Studies to find the optimal geometry and string spacing are currently underway.
Some of the geometries can be seen in Figure \ref{fig:gen2-geometries}.
All of the designs add 120 strings to the detector within the region of the South Pole station designated as the Dark Sector.
Uniform string spacings of 200\,m, 240\,m, and 300\,m, which instrument volumes of $6.0\,\mathrm{km}^{3}$, $8.0\,\mathrm{km}^{3}$, and $11.9\,\mathrm{km}^3$ respectively, have been studied.
Alternative array designs are also under study.
In addition, IceCube--Gen2's reach may further be enhanced by exploiting the air-shower detection and vetoing capabilities of an extended surface array, and by including an extended $100\,\mathrm{km}^2$ radio array to achieve improved sensitivity to neutrinos in the $10^{16}$--$10^{20}$ eV energy range, including cosmogenic neutrinos.

While the design details remain to be finalized,
IceCube--Gen2 will reveal an unobstructed view of the universe at PeV energies
where most of the universe is opaque to high-energy photons.
It will operate simultaneously with the next generation of electromagnetic and gravitational wave detectors, allowing for more multimessenger analyses.
With its unprecedented sensitivity
and improved angular resolution, this observatory
will enable detailed spectral studies as well as potential 
point source detections and other new discoveries.

\end{document}